\documentclass[sigconf,manuscript]{acmartcc}

\makeatletter                   
\def\mdseries@tt{m}             
\makeatother                    

\usepackage[plain]{fancyref}
\usepackage[draft=true]{minted} 
\usepackage{color}
\usepackage{textgreek}
\usepackage{breakurl}           

\usepackage{amssymb}
\usepackage{lipsum}
\usepackage[final]{pdfpages}
\usepackage{setspace}
\usepackage{epsf}
\usepackage{epsfig}
\usepackage{graphicx}
\usepackage{algorithmic}
\usepackage{amsmath}
\usepackage{letltxmacro}

\usepackage{enumitem}
\usepackage{tikz}
\usepackage{xspace}

\usepackage{url}
\usepackage{multirow}

  \renewenvironment{thebibliography}[1]{%
    \begin{oldthebibliography}{#1}%
      \setlength{\parskip}{0ex}%
      \setlength{\itemsep}{0ex}%
  }%
  {%
    \end{oldthebibliography}%
  }

\newwrite\arxivdeps
\immediate\openout\arxivdeps=\jobname-arxivdeps.log

\newcommand\verifymarkedforarxivfile[1]{%
\ifdefined\arxivbuild
\else
\IfFileExists{#1}%
{}%
{\GenericWarning{Marked file (#1) for inclusion in arxiv build does not exist}}
\fi%
}

\newcommand\markforarxiv[1]{%
\verifymarkedforarxivfile{#1}%
\write\arxivdeps{IncludeInArxiv: #1}%
}

\DeclareUrlCommand\UScore{\urlstyle{rm}}

\LetLtxMacro\oldincludegraphics\includegraphics
\renewcommand{\includegraphics}[2][]{%
\markforarxiv{#2}%
\oldincludegraphics[#1]{#2}}

\LetLtxMacro\oldincludepdf\includepdf
\renewcommand{\includepdf}[2][]{%
\markforarxiv{#2}%
\oldincludepdf[#1]{#2}}

\def\nfigure[#1,#2,#3]{
\begin{figure}
\vspace*{0mm}
\begin{center}

\includegraphics[width=\columnwidth]{#1} 
\vspace*{-6mm}\caption[]{#2
} \label{#3}

\vspace*{-3mm}
\end{center}
\end{figure}}

\def\cfigure[#1,#2,#3]{
\begin{figure}
\vspace*{0mm}
\begin{center}

\includegraphics[width=3in]{#1} 
\vspace*{-3mm}\caption[]{#2
} \label{#3}
 
\vspace*{-5mm}
\end{center}
\end{figure}}

\def\cfigurefour[#1,#2,#3]{
\begin{figure}
\vspace*{0mm}
\begin{center}

\includegraphics[width=4in]{#1} 
\vspace*{-3mm}\caption[]{#2
} \label{#3}
 
\vspace*{-5mm}
\end{center}
\end{figure}}

\def\cfiguretemp[#1,#2,#3]{
\begin{figure}
\vspace*{0mm}
\begin{center}

\includegraphics[width=3.5in]{#1} 
\vspace*{-3mm}\caption[]{#2
} \label{#3}
 
\vspace*{-5mm}
\end{center}
\vspace*{-2mm}
\end{figure}}

\def\wfigure[#1,#2,#3]{
\begin{figure*}
\vspace*{0mm}
\begin{center}
 \includegraphics[width=\textwidth]{#1} 
 \vspace*{-3mm}\caption[]{#2
} \label{#3}
 
\end{center}
\end{figure*}}

\def\threefigure[#1,#2,#3,#4,#5]{
\begin{figure*}
\vspace*{0mm}
\begin{center}

\begin{tabular}{ccc}
\includegraphics[width=2in]{#1} & \includegraphics[width=2in]{#2} &  \includegraphics[width=2in]{#3} \\
(a) & (b) & (c) \\
\end{tabular}
\vspace*{-3mm}\caption[]{#4
} \label{#5}

\vspace*{-5mm}
\end{center}
\vspace*{-2mm}
\end{figure*}}

\def\dcfigure[#1,#2,#3,#4,#5,#6]{
{
\begin{figure*}
\begin{center}
\begin{minipage}[c]{\columnwidth}{
\includegraphics[width=\columnwidth]{#1} 
\vspace*{0mm}\caption[]{#2} \label{#3} \
}\end{minipage}\hspace*{\columnsep}\
\begin{minipage}[c]{\columnwidth}{
\includegraphics[width=\columnwidth]{#4} 
\vspace*{0mm}\caption[]{#5}\label{#6} \
}\end{minipage}
\end{center}
\end{figure*}
}
}

\def\scfigure[#1,#2,#3]{
{
\begin{figure*}
\begin{center}
\begin{minipage}[c]{3.5in}{
\includegraphics[width=3.5in]{#1} 
}\end{minipage}
\caption[]{#2} \label{#3} \
\end{center}
\end{figure*}
}
}

\def\tableByTable[#1,#2,#3,#4,#5,#6]{
{
\begin{table*}
\begin{center}
\begin{minipage}[c]{3in}{
\centering
{#1}
\vspace*{0mm}\tabcaption[]{#2}\label{#3} \
}\end{minipage}\hspace*{\columnsep}\
\begin{minipage}[c]{3in}{
\centering
{#4}
\vspace*{0mm}\tabcaption[]{#5}\label{#6} \
}\end{minipage}
\end{center}
\end{table*}
}
}

\def\figureByTable[#1,#2,#3,#4,#5,#6]{
{
\begin{figure*}
\begin{center}
\begin{minipage}[c]{3in}{
\centering
\includegraphics[width=\textwidth]{#1}
\vspace*{0mm}\figcaption[]{#2} \label{#3} \
}\end{minipage}\hspace*{\columnsep}\
\begin{minipage}[c]{3.3in}{
\centering
{#4}
\vspace*{0mm}\tabcaption[]{#5}\label{#6} \
}\end{minipage}
\end{center}
\end{figure*}
}
}

\def\tableByFigure[#1,#2,#3,#4,#5,#6]{
{
\begin{figure*}
\begin{center}
\begin{minipage}[c]{4.3in}{
\centering
{#1}
\vspace*{0mm}\tabcaption[]{#2} \label{#3} \
}\end{minipage}\hspace*{\columnsep}\
\begin{minipage}[c]{2.2in}{
\centering
\includegraphics[width=\textwidth]{#4}
\vspace*{-0.35in}\caption[]{#5}\label{#6} \
}\end{minipage}
\end{center}
\end{figure*}
}
}

\def\doublecfigure[#1,#2,#3,#4]{
{
\begin{figure}
\begin{center}
\begin{minipage}[c]{1.5in}{
\begin{center}
\includegraphics[width=1.5in]{#1}
\end{center}
}\end{minipage}\hspace*{1em}\
\begin{minipage}[c]{1.5in}{
\begin{center}
\includegraphics[width=1.5in]{#2}
\end{center}
}\end{minipage}
\vspace*{0mm}\caption[]{#3} \label{#4} \
\end{center}
\end{figure}
}
}

\def\qcfigure[#1,#2,#3,#4,#5,#6]{
{
\begin{figure*}
\vspace*{0.2in}\
\begin{center}
\begin{minipage}[c]{3in}{
\includegraphics[width=3in]{#1} 
\vspace*{-3mm}
}
\end{minipage}\hspace*{0.5in}\
\begin{minipage}[c]{3in}{
\includegraphics[width=3in]{#2} 
\vspace*{-3mm}
}\end{minipage}

\begin{minipage}[c]{3in}{
\includegraphics[width=3in]{#3} 
\vspace*{-3mm}
}
\end{minipage}\hspace*{0.5in}\
\begin{minipage}[c]{3in}{
\includegraphics[width=3in]{#4} 
\vspace*{-3mm}
}\end{minipage}
\end{center}
\caption[]{#5}\label{#6}
\end{figure*}
}
}

\def\twfigure[#1,#2,#3,#4,#5]{
{
\begin{figure*}
\vspace*{0.2in}\
\begin{center}
\begin{minipage}[c]{6.5in}{
\includegraphics[width=6.5in]{#1} 
\vspace*{-3mm}
}
\end{minipage}

\begin{minipage}[c]{6.5in}{
\includegraphics[width=6.5in]{#2} 
\vspace*{-3mm}
}\end{minipage}

\begin{minipage}[c]{6.5in}{
\includegraphics[width=6.5in]{#3} 
\vspace*{-3mm}
}
\end{minipage}
\end{center}
\caption[]{#4}\label{#5}
\end{figure*}
}
}

\def\dwfigure[#1,#2,#3,#4]{
{
\begin{figure*}
\vspace*{0.2in}\
\begin{center}
\begin{minipage}[c]{6.5in}{
\includegraphics[width=6.5in]{#1} 
\vspace*{-3mm}
}
\end{minipage}

\begin{minipage}[c]{6.5in}{
\includegraphics[width=6.5in]{#2} 
\vspace*{-3mm}
}\end{minipage}

\end{center}
\caption[]{#3}\label{#4}
\end{figure*}
}
}

\def\dssfigure[#1,#2,#3,#4,#5,#6]{
{
\begin{figure*}
\vspace*{0.2in}\
\begin{center}
\begin{minipage}[c]{4in}{
\includegraphics[width=4in]{#1}
\vspace*{-3mm}\caption[]{#2} \label{#3} \
}\end{minipage}\hspace*{0.5in}\
\begin{minipage}[c]{2in}{
\includegraphics[width=2in]{#4}
\vspace*{-3mm}\caption[]{#5}\label{#6} \
}\end{minipage}
\end{center}
\vspace*{-0.4in}\
\end{figure*}
}
}

\def\dsfigure[#1,#2,#3,#4,#5,#6]{
{
\begin{figure*}
\vspace*{0.2in}\
\begin{center}
\begin{minipage}[c]{3in}{
\includegraphics[width=3in]{#1}
\vspace*{-3mm}\caption[]{#2} \label{#3} \
}\end{minipage}\hspace*{0.5in}\
\begin{minipage}[c]{3in}{
\hspace*{0.5in}\
\includegraphics[height=3in]{#4}
\vspace*{-3mm}\caption[]{#5}\label{#6} \
}\end{minipage}
\end{center}
\vspace*{-0.4in}\
\end{figure*}
}
}

\def\dsyfigure[#1,#2,#3,#4,#5,#6]{
{
\begin{figure*}
\vspace*{0.2in}\
\begin{center}
\begin{minipage}[c]{2.5in}{
\includegraphics[height=2.5in]{#1}
\vspace*{-3mm}\caption[]{#2} \label{#3} \
}\end{minipage}\hspace*{0.5in}\
\begin{minipage}[c]{2.5in}{
\includegraphics[height=2.5in]{#4}
\vspace*{-3mm}\caption[]{#5}\label{#6} \
}\end{minipage}
\end{center}
\vspace*{-0.4in}\
\end{figure*}
}
}

\def\dyfigure[#1,#2,#3,#4,#5,#6]{
{
\begin{figure*}
\vspace*{0.2in}\
\begin{center}
\begin{minipage}[c]{3in}{
\includegraphics[height=3in]{#1} 
\vspace*{-3mm}\caption[]{#2} \label{#3} \
}\end{minipage}\hspace*{0.5in}\
\begin{minipage}[c]{3in}{
\includegraphics[height=3in]{#4} 
\vspace*{-3mm}\caption[]{#5}\label{#6} \
}\end{minipage}
\end{center}
\vspace*{-0.4in}\
\end{figure*}
}
}

\def\dyoldfigure[#1,#2,#3,#4,#5,#6]{
{
\begin{figure*}
\vspace*{0.2in}\
\begin{center}
\begin{minipage}[c]{3in}{
\epsfysize=2.0in\
\hspace{0.5in}\
\epsfbox{#1}
\vspace*{-3mm}\caption[]{#2} \label{#3} \
}\end{minipage}\hspace*{0.25in}\
\begin{minipage}[c]{3in}{
\epsfysize=2.0in\
\hspace{0.5in}\
\epsfbox{#4}
\vspace*{-3mm}\caption[]{#5}\label{#6} \
}\end{minipage}
\end{center}
\vspace*{-0.4in}\
\end{figure*}
}
}

\def\cfiguredouble[#1,#2,#3,#4]{
\begin{figure}
\vspace*{0.2in}\
\begin{center}
\begin{minipage}[c]{1.5in}{
\epsfxsize=1.5in\
\epsfbox{#1}
}\end{minipage}\hspace*{0.1in}\
\begin{minipage}[c]{1.5in}{
\epsfxsize=1.5in\
\vspace{0.1in}\epsfbox{#2}
}\end{minipage}\vspace*{-0.10in} \caption[]{#3}\label{#4}
\end{center}
\vspace*{-0.4in}\
\end{figure}
}

\def\wpfigure[#1,#2,#3,#4]{
\begin{figure*}
\vspace*{4mm}
\begin{center}

\includegraphics[width=#4]{#1} 

\vspace*{-3mm}\caption[]{#2
} \label{#3}

\vspace*{-5mm}
\end{center}
\end{figure*}}

\def\wprfigure[#1,#2,#3,#4,#5]{
\begin{figure*}
\vspace*{4mm}
\begin{center}

\includegraphics[width=#4, angle=#5]{#1} 

\vspace*{-3mm}\caption[]{#2
} \label{#3}

\vspace*{-5mm}
\end{center}
\end{figure*}}

\def\DoubleFigureWSlide[#1,#2,#3,#4,#5,#6,#7,#8,#9]{
\begin{figure*}
\vspace*{#9}
\begin{center}
\begin{minipage}{#4}
\includegraphics[width=#4]{#1}
\vspace*{-3mm}\caption{#2
}\label{#3}
\end{minipage}
\hspace{2em}
\begin{minipage}{#8}
\includegraphics[width=#8]{#5}
\vspace*{-3mm}\caption{#6
}\label{#7}
\end{minipage}
\vspace*{-5mm}
\end{center}
\end{figure*}
}

\def\DoubleFigureW[#1,#2,#3,#4,#5,#6,#7,#8]{
\begin{figure*}
\vspace*{0in}
\begin{center}
\begin{minipage}{#4}
\includegraphics[width=#4]{#1}
\vspace*{-3mm}\caption{#2
}\label{#3}
\end{minipage}
\hspace{2em}
\begin{minipage}{#8}
\includegraphics[width=#8]{#5}
\vspace*{-3mm}\caption{#6
}\label{#7}
\end{minipage}
\vspace*{-5mm}
\end{center}
\end{figure*}
}

\def\DoubleFigureWHack[#1,#2,#3,#4,#5,#6,#7,#8]{
\begin{figure*}
\vspace*{0in}
\begin{center}
\begin{minipage}{3in}
\includegraphics[width=#4]{#1}
\vspace*{-3mm}\caption{#2
}\label{#3}
\end{minipage}
\hspace{2em}
\begin{minipage}{3in}
\includegraphics[width=#8]{#5}
\vspace*{-3mm}\caption{#6
}\label{#7}
\end{minipage}
\vspace*{-5mm}
\end{center}
\end{figure*}
}

\def\ddcfigure[#1,#2,#3,#4]{
\begin{figure*}
\vspace*{0.2in}\
\begin{center}
\begin{minipage}[c]{\columnwidth}{
\includegraphics[width=\columnwidth]{#1} 
}\end{minipage}\hspace{0.5in}\
\begin{minipage}[c]{\columnwidth}{
\includegraphics[width=\columnwidth]{#2} 
}\end{minipage} \caption[]{#3}\label{#4}
\end{center}
\end{figure*}
}

\def\ddcfigureSlide[#1,#2,#3,#4,#5]{
\begin{figure*}
\vspace*{#5}\
\begin{center}
\begin{minipage}[c]{3in}{
\includegraphics[height=3in]{#1} 
}\end{minipage}\hspace{0.5in}\
\begin{minipage}[c]{3in}{
\includegraphics[height=3in]{#2} 
}\end{minipage}\vspace*{-0.10in} \caption[]{#3}\label{#4}
\end{center}
\vspace*{-0.4in}\
\end{figure*}
}

\def\cxfigure[#1,#2,#3]{
\begin{figure}
\vspace*{4mm}
\begin{center}
 
\epsfxsize=2.5in\
\epsfbox{#1}\
 
\vspace*{-0.10in}\caption[]{#2
} \label{#3}
 
\vspace*{-5mm}
\end{center}
\vspace*{-2mm}
\end{figure}}

\newif\ifremark
\long\def\remark#1{
        \begingroup%
        \dimen0=\columnwidth
        \advance\dimen0 by -1in%
        \setbox0=\hbox{\parbox[b]{\dimen0}{\protect\em #1}}
        \dimen1=\ht0\advance\dimen1 by 2pt%
        \dimen2=\dp0\advance\dimen2 by 2pt%
        \vskip 0.25pt%
        \hbox to \columnwidth{%
                \vrule height\dimen1 width 3pt depth\dimen2%
                \hss\copy0\hss%
                \vrule height\dimen1 width 3pt depth\dimen2%
        }%
        \endgroup%
}

\usepackage{xcolor}

\definecolor{cyanish}{rgb}{0,0.8,1.0}
\definecolor{orange}{rgb}{1.0,0.5,0.0}
\definecolor{pink}{rgb}{1.0,0.47,0.6}
\definecolor{light-gray}{gray}{0.95}
\definecolor{jiancolor}{RGB}{0,153,153}
\definecolor{mygreen}{RGB}{50,200,50}
\definecolor{pink}{rgb}{1.0,0.47,0.6}

\definecolor{commentgreen}{rgb}{0.0,0.5,0.0}

\newcommand{\ignore}[1]{}

\newcommand{\reflns}[2]{Lines~\hyperref[#1]{\ref*{#1}-\ref*{#2}}}

\newcommand{\x}[1]{$\times$}

\newif\ifcutforspace
\long\def\cutforspace#1{
\ifcutforspace%
        \begingroup%
        \dimen0=\columnwidth
        \advance\dimen0 by -1in%
        \setbox0=\hbox{\parbox[b]{\dimen0}{\protect{\em Cut For Space} #1}}
        \dimen1=\ht0\advance\dimen1 by 2pt%
        \dimen2=\dp0\advance\dimen2 by 2pt%
        \vskip 0.25pt%
        \hbox to \columnwidth{%
                \vrule height\dimen1 width 3pt depth\dimen2%
                \hss\copy0\hss%
                \vrule height\dimen1 width 3pt depth\dimen2%
        }%
        \endgroup%
\fi}

\newcommand{\nova}[1]{NOVA} 

\usepackage{csquotes}

\newcommand{\Mats}[1]{Mats}
\newcommand{\typedschematicblock}[1]{typed schematic block}
\newcommand{\Typedschematicblock}[1]{Typed schematic block}
\newcommand{\TypedSchematicBlock}[1]{Typed Schematic Block}
\newcommand{\interfacetype}[1]{interface type}
\newcommand{\Interfacetype}[1]{Interface type}
\newcommand{\InterfaceType}[1]{Interface Type}
\newcommand{\interfacecheking}[1]{interface checking}
\newcommand{\Interfacecheking}[1]{Interface checking}
\newcommand{\InterfaceChecking}[1]{Interface Checking}
\newcommand{\interactiveblock}[1]{interactive block}
\newcommand{\Interactiveblock}[1]{Interactive block}
\newcommand{\InteractiveBlock}[1]{Interactive Block}

\def\paperFigure[#1,#2,#3,#4,#5]{
    \begin{figure}
        \centering
        \includegraphics[#1]{#2}
        \caption[]{#3} \label{#4}
        \Description{#5}
    \end{figure}}

\def\paperFigureWide[#1,#2,#3,#4,#5]{
    \begin{figure*} 
        \centering
        \includegraphics[#1]{#2}
        \caption[]{#3} \label{#4}
        \Description{#5}
    \end{figure*}}

\AtBeginDocument{%
  \providecommand\BibTeX{{%
    \normalfont B\kern-0.5em{\scshape i\kern-0.25em b}\kern-0.8em\TeX}}}

\setcopyright{cc}
\copyrightyear{2023}
\acmYear{2023}
\acmDOI{https://arxiv.org/abs/2303.06620}

\acmConference[CHI '23]{CHI '23: Workshop on electronic prototyping tools and toolkits}{April 23--28, 2023}{Hamburg, Germany}
\acmBooktitle{CHI '23 Workshop [WS2]: Beyond Prototyping Boards: Future Paradigms for Electronics Toolkits,
  April 23--28, 2023, Hamburg, Germany}
\acmPrice{15.00}
\acmISBN{978-1-4503-XXXX-X/18/06}

\begin{document}

\title{PCB-ready breakout boards: Bridging the gap between electronics prototyping and production}

\author{J. Garza}
\affiliation{%
  \department{Computer Science and Engineering}
  \institution{UC San Diego}
  \city{San Diego}
  \state{CA}
  \country{USA}
}
\email{jgarzagu@eng.ucsd.edu}



\author{Steven Swanson}
\affiliation{%
  \department{Computer Science and Engineering}
  \institution{UC San Diego}
  \city{San Diego}
  \state{CA}
  \country{USA}
}
\email{swanson@eng.ucsd.edu}
\date{}

\renewcommand{\shortauthors}{Garza, et al.}



\begin{abstract}
    Electronics prototyping using breakout boards allows designers with and without an engineering background to rapidly create interactive prototypes. However, when it comes to transition to a production-ready PCB design, stagnation exists due to the high skill floor required for PCB design. While PCB design automation has been used successfully in recent research tools to reduce the required expertise, little has been done to integrate these tools directly into the electronics prototyping cycle. This position paper aims to bring attention to the possibility of integrating recent PCB design automation paradigms into the electronics prototyping cycle for the creation of PCB-ready breakout boards: breakout boards whose designs would have the ability to be pipelined directly into new user interfaces that leverage the use of automation for the rapid creation of production-ready PCB designs.
\end{abstract}

\maketitle

\section{Background}
\label{sec:background}

Prototyping using breakout boards is one of the three most popular paradigms when it comes to prototyping with electronic components~\cite{lambrichts2021survey}. Breakout boards consist of a set of modular PCBs, each containing multiple electronic components, which can be easily interconnected using breadboards for interactive circuit design exploration or debugging. Anyone can design and distribute breakout boards without affecting the compatibility between them, popular distributors include Arduino~\cite{arduinoref}, Sparkfun~\cite{Sparkfun}, and Adafruit~\cite{adafruit}.

After circuit prototyping, designers transfer the physical connections made between the breakout boards into a virtual representation format known as an electronic schematic. Electronic schematics provide designers with a simpler visual interactive format for connecting electronic components, where even dimensional blocks are used instead of irregular electronic component footprint shapes. After the schematic design, designers create a PCB layout by drawing virtual copper trace lines between the footprints of the electronic components. Finally, after generating the schematic and PCB layout of the circuit, the circuit design is considered ready for manufacturing. Software tools that assist designers in creating schematics and PCB layouts are known as PCB design tools, examples include KiCad~\cite{kiCAD}, Altium~\cite{Altium}, Fusion 360~\cite{EAGLE}. 

Breakout board prototyping and PCB design require different techniques and design skills and are therefore considered separate stages, being PCB design exponentially more difficult for beginners, making this transition unachievable for many. However we believe there are new opportunities to bring breakout board prototyping closer to PCB design. In this position paper, we explain the challenges of transition between breakout boards and PCB design, we propose our vision and differentiate it from related work, mention some research opportunities, and end with initial work in this direction.

\section{Translating breakout board design prototypes to PCB designs}

Moving from an electronic prototype based on breakout boards to a PCB design is, in most cases, a desired step and often influences the prototyping platforms that are chosen. A recent user study shows that the ability of prototyping platforms to evolve to a custom PCB is the third most important feature for the electronics engineer, in addition to the platform's ease of use and the ability to make changes quickly~\cite{lambrichts2021survey}. While the ease of translating a prototype design to a PCB design is desirable, it is not easy, and to improve the translation process it is first necessary to identify the challenges.

Through the teaching of various courses where students with no previous experience in electronics learn circuit design, we have identified some of the challenges when translating electronic prototypes using breakout boards to PCB designs, which can be summarized as follows: (1) A steep learning curve required to use PCB design tools (e.g.,Altium, Fusion 360, KiCad), (2) Understanding each of the components on the breakout boards and separating those required for the final design.

While (1) is clear (2) is not a well-studied problem. In addition to learning PCB design tools, after electronics prototyping, designers have to figure out how to correctly transfer the designs from each breakout board in the prototype design to a PCB design. This involves separating non-important components within the breakout boards (e.g., power regulators, optional configurations, etc.), reviewing all data sheets, redrawing the schematics and PCB layouts, and merging designs. A process that is often complex for beginners and tedious for experts. As an analogy, this would be like integrating a code library into a programming code by going through each method manually and extracting the parts of the code to use. In this process, we believe that there are alternative solutions.

\section{PCB-ready breakout boards vision}
\label{sec:overview}

In the future, we anticipate that breakout boards will be ready to be incorporated into PCB design automation tools that can automatically generate the schematic and PCB layout of prototype designs with minimal effort. Reducing the gap between electronic prototyping and PCB design. This will allow designers, engineers, makers and researchers to easily transition from prototype ideas to products. We believe that the impact of these types of tools would be significant. Making more ideas that normally remain in prototypes to electronic designs closer to commercial products, thereby bringing more innovation to society. 

For this idea to be feasible, subtle design changes must be made to the schematics that ship with the breakout boards in order to integrate them into automated PCB design tools. We call breakout boards that are ready to be integrated into PCB layout automation tools as \textbf{\textit{PCB-ready breakout boards}}. 

PCB-ready breakout boards designs will be incorporated into new user interfaces to automatically generate complete production-ready PCB layouts, instead of the traditional consuming process of copying and redesigning each breakout board design to be incorporated into the circuit design. These new interactive interfaces will range from a simple selection interface of breakout boards to use to more advanced interfaces that allow for multiple design configurations (e.g., I/O pin selections, design constraint validations, part replacements, etc.). 

Several existing approaches have realized part but not all of this vision. Commercial tools such as Sparkfun's À La Carte (ALC)~\cite{SparkfunALC} and Altium's Upverter Modular~\cite{AltiumUpverterModular} have taken small steps to bridge the gap between prototype and production. These tools allow designers to select different breakout boards, development boards or pre-tested circuit blocks, and allow their integration to create complete PCB designs. However these tools are non-automated, and require one or more engineers to manually create the full schematic and PCB layout design. A process with a design fee that can approach \$1,000 per design~\cite{SparkfunALCFaq}. In addition, not having standards for scaling the number of usable modules prevents community collaboration, limiting the designs that can be achieved with these tools.

Compared to Sparkfun ALC and Altium Upverter, the tools we envision will be fully automated, with no additional engineering services required, allowing for rapid design iterations. Also, if the process is standardized, it will allow for community collaboration, without limiting the number of designs to a few. 

Recent research tools have explored new circuit design approaches that take advantage of automation to rapidly create schematic and PCB layout designs. Echidna~\cite{merrill2019echidna}, Appliancizer~\cite{garza2021appliancizer}, Polymorphic Blocks~\cite{lin2020polymorphic} are examples of PCB design automation tools that leverage automation to explore new design paradigms with the goal of facilitating design by increasing design abstraction. Echidna enables the automatic generation of ready-to-manufacturer PCBs from virtual modular schematic designs. In Echidna, the design is abstracted to the selection of components to be used, with that information the software generates a complete PCB layout output. Appliancizer, goes further, abstracting the design to the GUI level, allowing automated PCB layout, firmware, and interface simulation generation from web interfaces. Polymorphic Blocks takes a different approach by allowing automated generation of schematic designs from hardware description languages.

While these research tools allow for automated generation of schematics and PCB layouts from higher levels of abstraction, these tools are not integrated into the physical electronic prototyping process cycle. However, similarities exist between breakout boards and the research tools described above. Both approaches use a modular or blocks design approach for ease of design. While breakout boards present modular circuit designs in a physical format, PCB layout automation research tools use libraries containing virtual modular circuit designs for design automation. These similarities make integration between the two design cycles not a far off possibility.

In contrast, Echidna, Appliancizer, and Polymorphic Blocks, whose libraries and module design are virtual and therefore cannot be easily tested separately, PCB-ready breakout boards will enable both: a physical exploration of designs and a rapid transition to production-ready electronic designs. Furthermore, we envision the encapsulation of information within breakout board designs to be easy and scalable and not something that requires in-depth knowledge (e.g., learning new programming languages, building large databases) as in previous research tools.

\section{Research opportunities}

In this vision, there are still many research questions to be answered, some of which are the following:

\begin{itemize}
    \item What is the optimal way to include additional parameters into breakout board designs for use by PCB design automation tools?
    \item How to design the user interfaces to include multiple settings and at the same time be easy to use?
    \item How to allow multiple design configurations required for each module (e.g., adding or removing resistors to set I2C address)?
    \item How to include validations so that the generated PCB design is safe of design errors?
\end{itemize}

We also see necessary a user study to correctly identify transition difficulties between prototyping and PCB design.

\subsection{Further challenges and opportunities}

With this vision we also anticipate possible challenges that the tools may have, some of which are the following.

\textit{Obsolete Electronic Components} – In the proposed system, the components used on the breakout boards will be the same as those used in the final PCB design. Faced with an obsolete electronic component, the design would not be possible. Methods of replacing obsolete electronic components or rapidly updating breakout boards designs are necessary.

\textit{PCB Optimizations} – In the proposed system, breakout board PCB layouts would be automatically copied and pasted, and the missing nets would be routed automatically for ease of design. In this approach, the final PCB layout would lack potential routing optimizations compared to a manual layout design. However, we see that in the future PCB auto-routing improvements could efficiently layout the complete PCB.

\section{Commercialization opportunities}

Breakout boards or integrated toolkits that adopt a PCB-ready design might become more attractive to users compared to standard breakouts boards, as the user has the added benefit of ease of translation from prototype to product. 
\section{Work in this direction}
\label{sec:system}

Our team has started taking steps in this direction towards PCB-ready breakout boards. Our current system allows annotations on existing schematics to create schematic blocks. These schematic blocks can be imported into a simplified user interface that validates the connections and generates a complete schematic. The next step is to generate the PCB layout from blocks which has already been demonstrated in previous tools.

\section{RESEARCH TEAM}
\label{sec:conclude}

\hspace{\parindent}\textbf{J. Garza} is a Ph.D. student at University of California, San Diego, where he explores new methods to facilitate and accelerate the design of electronic devices by creating computer-aided design (CAD) tools with new workflows and interactions.

\indent\textbf{Steven Swanson} is a professor in the Department of Computer Science and Engineering at the University of California, San Diego and the director of the Non-volatile Systems Laboratory. He has received an NSF CAREER Award, Google Faculty Awards,  a Facebook Faculty Award, and been a NetApp Faculty Fellow.  He is a co-founder of the Non-Volatile Memories Workshop.


\bibliography{paper.bib}


\begin{thebibliography}{13}


\ifx \showCODEN    \undefined \def \showCODEN     #1{\unskip}     \fi
\ifx \showDOI      \undefined \def \showDOI       #1{#1}\fi
\ifx \showISBNx    \undefined \def \showISBNx     #1{\unskip}     \fi
\ifx \showISBNxiii \undefined \def \showISBNxiii  #1{\unskip}     \fi
\ifx \showISSN     \undefined \def \showISSN      #1{\unskip}     \fi
\ifx \showLCCN     \undefined \def \showLCCN      #1{\unskip}     \fi
\ifx \shownote     \undefined \def \shownote      #1{#1}          \fi
\ifx \showarticletitle \undefined \def \showarticletitle #1{#1}   \fi
\ifx \showURL      \undefined \def \showURL       {\relax}        \fi
\providecommand\bibfield[2]{#2}
\providecommand\bibinfo[2]{#2}
\providecommand\natexlab[1]{#1}
\providecommand\showeprint[2][]{arXiv:#2}

\bibitem[Adafruit(2023)]%
        {adafruit}
\bibfield{author}{\bibinfo{person}{Adafruit}.} \bibinfo{year}{2023}\natexlab{}.
\newblock \bibinfo{title}{Adafruit Official Website - DIY Electronics and
  Kits}.
\newblock
\newblock
\urldef\tempurl%
\url{https://www.adafruit.com/}
\showURL{%
\tempurl}


\bibitem[Altium(2023a)]%
        {Altium}
\bibfield{author}{\bibinfo{person}{Altium}.} \bibinfo{year}{2023}\natexlab{a}.
\newblock \bibinfo{title}{Altium Designer. (2023)}.
\newblock
\newblock
\urldef\tempurl%
\url{https://www.altium.com/altium-designer/}
\showURL{%
\tempurl}


\bibitem[Altium(2023b)]%
        {AltiumUpverterModular}
\bibfield{author}{\bibinfo{person}{Altium}.} \bibinfo{year}{2023}\natexlab{b}.
\newblock \bibinfo{title}{Upverter Modular (2023)}.
\newblock
\newblock
\urldef\tempurl%
\url{https://upverter.com/}
\showURL{%
\tempurl}


\bibitem[Arduino(2020)]%
        {arduinoref}
\bibfield{author}{\bibinfo{person}{Arduino}.} \bibinfo{year}{2020}\natexlab{}.
\newblock \bibinfo{title}{Arduino Reference}.
\newblock
\newblock
\urldef\tempurl%
\url{https://www.arduino.cc/reference/en/}
\showURL{%
\tempurl}


\bibitem[EAGLE(2020)]%
        {EAGLE}
\bibfield{author}{\bibinfo{person}{EAGLE}.} \bibinfo{year}{2020}\natexlab{}.
\newblock \bibinfo{title}{PCB Design Software - Autodesk}.
\newblock
\newblock
\urldef\tempurl%
\url{https://www.autodesk.com/products/eagle/overview}
\showURL{%
\tempurl}


\bibitem[Electronics(2023a)]%
        {Sparkfun}
\bibfield{author}{\bibinfo{person}{SparkFun Electronics}.}
  \bibinfo{year}{2023}\natexlab{a}.
\newblock \bibinfo{title}{SparkFun}.
\newblock
\newblock
\urldef\tempurl%
\url{https://www.sparkfun.com/}
\showURL{%
\tempurl}


\bibitem[Electronics(2023b)]%
        {SparkfunALC}
\bibfield{author}{\bibinfo{person}{SparkFun Electronics}.}
  \bibinfo{year}{2023}\natexlab{b}.
\newblock \bibinfo{title}{SparkFun À La Carte. (2022)}.
\newblock
\newblock
\urldef\tempurl%
\url{https://alc.sparkfun.com}
\showURL{%
\tempurl}


\bibitem[Electronics(2023c)]%
        {SparkfunALCFaq}
\bibfield{author}{\bibinfo{person}{SparkFun Electronics}.}
  \bibinfo{year}{2023}\natexlab{c}.
\newblock \bibinfo{title}{SparkFun À La Carte Faq. (2022)}.
\newblock
\newblock
\urldef\tempurl%
\url{https://alc.sparkfun.com/faq}
\showURL{%
\tempurl}


\bibitem[Garza et~al\mbox{.}(2021)]%
        {garza2021appliancizer}
\bibfield{author}{\bibinfo{person}{Jorge Garza}, \bibinfo{person}{Devon~J
  Merrill}, {and} \bibinfo{person}{Steven Swanson}.}
  \bibinfo{year}{2021}\natexlab{}.
\newblock \showarticletitle{Appliancizer: Transforming Web Pages into
  Electronic Devices}. In \bibinfo{booktitle}{\emph{Proceedings of the 2021 CHI
  Conference on Human Factors in Computing Systems}}. \bibinfo{pages}{1--13}.
\newblock


\bibitem[KiCad(2022)]%
        {kiCAD}
\bibfield{author}{\bibinfo{person}{KiCad}.} \bibinfo{year}{2022}\natexlab{}.
\newblock \bibinfo{title}{KiCad EDA}.
\newblock
\newblock
\urldef\tempurl%
\url{https://www.kicad.org/}
\showURL{%
\tempurl}


\bibitem[Lambrichts et~al\mbox{.}(2021)]%
        {lambrichts2021survey}
\bibfield{author}{\bibinfo{person}{Mannu Lambrichts}, \bibinfo{person}{Raf
  Ramakers}, \bibinfo{person}{Steve Hodges}, \bibinfo{person}{Sven Coppers},
  {and} \bibinfo{person}{James Devine}.} \bibinfo{year}{2021}\natexlab{}.
\newblock \showarticletitle{A survey and taxonomy of electronics toolkits for
  interactive and ubiquitous device prototyping}.
\newblock \bibinfo{journal}{\emph{Proceedings of the ACM on Interactive,
  Mobile, Wearable and Ubiquitous Technologies}} \bibinfo{volume}{5},
  \bibinfo{number}{2} (\bibinfo{year}{2021}), \bibinfo{pages}{1--24}.
\newblock


\bibitem[Lin et~al\mbox{.}(2020)]%
        {lin2020polymorphic}
\bibfield{author}{\bibinfo{person}{Richard Lin}, \bibinfo{person}{Rohit
  Ramesh}, \bibinfo{person}{Connie Chi}, \bibinfo{person}{Nikhil Jain},
  \bibinfo{person}{Ryan Nuqui}, \bibinfo{person}{Prabal Dutta}, {and}
  \bibinfo{person}{Bj{\"o}rn Hartmann}.} \bibinfo{year}{2020}\natexlab{}.
\newblock \showarticletitle{Polymorphic blocks: Unifying high-level
  specification and low-level control for circuit board design}. In
  \bibinfo{booktitle}{\emph{Proceedings of the 33rd Annual ACM Symposium on
  User Interface Software and Technology}}. \bibinfo{pages}{529--540}.
\newblock


\bibitem[Merrill et~al\mbox{.}(2019)]%
        {merrill2019echidna}
\bibfield{author}{\bibinfo{person}{Devon~J Merrill}, \bibinfo{person}{Jorge
  Garza}, {and} \bibinfo{person}{Steven Swanson}.}
  \bibinfo{year}{2019}\natexlab{}.
\newblock \showarticletitle{Echidna: mixed-domain computational implementation
  via decision trees}. In \bibinfo{booktitle}{\emph{Proceedings of the 3rd
  Annual ACM Symposium on Computational Fabrication}}. \bibinfo{pages}{1--12}.
\newblock


\end{thebibliography}
\bibliographystyle{ACM-Reference-Format}

\end{document}